\documentclass[preprint,superscriptaddress,nofootinbib]{revtex4}
\usepackage{graphicx}
\usepackage{braket}
\newcommand{\del}{\partial}

\begin{document}
\title{Torsional instanton effects in quantum gravity}

\author{Romesh K. Kaul}
\email{kaul@imsc.res.in}
\affiliation{Institute of Mathematical Sciences\\
Chennai-600113, INDIA.}

\author{Sandipan Sengupta}
\email{sandipan@iucaa.ernet.in}
\affiliation{Raman Research Institute\\
Bangalore-560080, INDIA\footnote{Address after 22nd Nov, 2014: Inter-University Centre for Astronomy and Astrophysics (IUCAA), Pune-411007, INDIA}.}


\begin{abstract}
We show that, in the first order gravity theory coupled to axions,
the instanton number of the Giddings-Strominger wormhole can
be interpreted as the Nieh-Yan topological index. The axion charge
of the baby universes is quantized in terms of the Nieh-Yan
integers. Tunneling between universes of different Nieh-Yan
charges implies a nonperturbative vacuum state. The associated
topological vacuum angle can be identified with the Barbero-Immirzi
parameter.
%
%
  %
\end{abstract}
  
\maketitle
\section{Introduction}
The study of topology change of spacetime due to quantum fluctuations dates back to Wheeler \cite{wheeler}, and continues to be a fascinating endeavour for physicists. While a complete and consistent quantum theory of gravity is still missing, one hopes that an analysis of the role of geometrical or topological fluctuations might bring out certain qualitative features of small scale gravity which are otherwise difficult to unravel. In fact it is this perspective from which the functional integral formulation of quantum gravity can still be thought to be relevant \cite{eguchi}, even though this theory is known to exhibit perturbative nonrenormalizability.

There exists a considerable amount of literature regarding gravitational instantons which are responsible for the change in topology of spacetime, both in pure and matter-coupled gravity theories. Among these, what we shall be concerned with in this article is the wormhole instanton \cite{hawking,stro,lav,gidd,gidd1,schl}. These are known to appear in the Euclidean theory of gravity coupled to antisymmetric tensor gauge field of rank two. Their existence as finite action solutions of axionic gravity was first noted by Giddings and Strominger \cite{gidd}. Whereas the wormhole itself acts as a bridge connecting two asymptotically flat universes, the semi-wormhole, which is just the half of it, can be interpreted as an instanton which leads to the creation or annihilation of a baby universe. Each such small universe has the topology of a three-sphere at any instant of time. This instanton effectively tunnels between two topologically distinct three geometries, namely, $R^3$ and $R^3+S^3$. Subsequently, these authors have used these configurations to study various low energy effects, e.g. loss of coherence and CP violation. Wormhole physics has also been applied by Coleman to argue that the cosmological constant should vanish and that coupling constants in general should get renormalized \cite{coleman,hawking1,kleb,wein}. 
That the effects of topology change due to a small wormhole can be represented by the insertion of an effective local vertex operator in the correlation functions has been observed in ref.\cite{jnan}.

In quantum field theory, any nontrivial instanton physics is generally associated with the emergence of a nonperturbative quantum vacuum which supercedes the naive perturbative vacuum state \cite{raja}. Such a nontrivial structure is essentially characterised by an additional free parameter in the theory, namely, a topological coupling constant. These nonperturbative effects can be ascribed solely to the existence of a topological density in the classical theory itself, which could be included as an additional term in an effective Lagrangian. Such a term will come multiplied by the same topological parameter which shows up as the vacuum angle in the quantum theory. Thus, even though topological densities, being total divergences, do not affect the classical dynamics, they do show up in the quantum theory. For example, in gauge theories such as the Quantum Chromodynamics (QCD), the origin of the famous $\theta$-vacuum lies in the fact that there exists a topological density in the theory, namely, the Pontryagin class, which can be added to the original Lagrangian alongwith a free coefficient $\theta$. Although classical physics remains insensitive to this addition, quantum theory perceives it and exhibits a nonperturbative structure of the vacuum. While the naive perturbative vacua are degenerate and are characterised by different values of the winding number, the instanton configurations break this degeneracy by tunneling between these vacua. As a result, the true vacuum state, also known as the $\theta$-vacuum, is a superposition of all the perturbative vacua weighted by $\theta$-dependent phases. 

Similar to gauge theories, gravity theory in four dimensions also admits an effective Lagrangian with additional terms which are topological densities. However, for gravity, there are three such terms, namely, Euler, Pontryagin and Nieh-Yan \cite{date,kaul,nieh,hehl}. Whereas the first two depend on the curvature tensor, the third depends only on torsion \cite{nieh}. In other words, Nieh-Yan invariant can be nontrivial only for configurations with a nonvanishing torsion. There exist many examples of gravitational instantons carrying nontrivial Euler and Pontryagin numbers \cite{hawking2,eguchi1,gibbons,smilga}. However, configurations that correspond to nonzero Nieh-Yan numbers as well as satisfy the equations of motion are not known to exist in pure gravity theory. Since torsion is identically zero in second order gravity, such configurations, if possible, can necessarily live only in the first order formulation (with or without matter) where tetrad and spin-connection are treated as independent fields \cite{tseytlin}. In view of this, axionic gravity, which admits a first order theory with nonvanishing torsion and exhibits instanton solutions, appears as a natural candidate where Nieh-Yan instantons can exist.

Also, keeping the general features of instanton physics in the case of gauge theories in mind, there is a reason to believe that the semi-wormhole instanton in axionic gravity must represent tunneling between topologically distinct `vacua'. The topological  numbers characterizing  these vacua   could be  associated with one of three possible  topological classes
in gravity theory, namely, Pontryagin, Nieh-Yan or Euler. However, such an interpretation has been missing so far. 

Here we provide a unified answer to both the issues raised above. First, we set up a first order formulation of gravity theory coupled to axionic matter, which admits the Giddings-Strominger wormholes as solutions. Then, we show that within this framework, the wormholes can be interpreted as torsional pseudoparticles, whose instanton number is the same as the Nieh-Yan index of the baby universe created (or annihilated) by the instanton.

Torsional instantons have also been discussed earlier in theories of gravity with or without matter \cite{regge,zanelli,auria,leigh}. Among the pure gravity configurations, the one found by Hanson and  Regge \cite{regge} is neither a solution of equations of motion nor does it carry a Nieh-Yan charge. On the other hand, the Chandia-Zanelli instantons \cite{zanelli}, which are not solutions of equations of motion, do carry nontrivial Nieh-Yan number. The torsion is purely geometric in these cases, having its origin in the degeneracy of tetrad. In contrast, in our case here, the torsion is generated by the antisymmetric tensor gauge field (or axion) through a special choice of its coupling to gravity.

In the next section, we present a first order action formulation for axionic gravity with nonvanishing torsion. This theory is then used to provide a new interpretation of the instanton number of the (semi-)wormholes. We also find that the axion charge is quantized in terms of Nieh-Yan integers. Next, we discuss the nontrivial vacuum structure originating due to wormhole effects which introduce an angular parameter $\eta$ as a new quantum coupling. This constant is   inverse of the Barbero-Immirzi parameter of Loop Quantum Gravity \cite{barbero}. The details of the large gauge transformations which induce a change of the Nieh-Yan number are also elaborated. Finally, we make a few concluding remarks.
  
\section{First order Lagrangian for axionic gravity:}
We study the following first order Lagrangian density for (Euclidean) gravity theory coupled to an antisymmetric tensor gauge field of rank two:
\begin{eqnarray}\label{LH} 
L(e,\omega,B)~=~-\frac{1}{2\kappa^2}e e^{\mu}_{I}e^{\nu}_{J}R_{\mu\nu}^{~IJ}(\omega)~+~\frac{1}{2\kappa}eH^{\mu\nu\alpha}e_{\mu}^I D_{\nu}(\omega) e_{\alpha I}~+~\beta e H^{\mu\nu\alpha}H_{\mu\nu\alpha}
\end{eqnarray} 
where $R_{\mu\nu}^{~IJ}(\omega)=\del_{[\mu}\omega_{\nu]}^{~IJ}+\omega_{[\mu}^{~IL}
\omega_{\nu]L}^{~~~J}$, $D_{\mu}(\omega)e_{\nu}^I=\del_{\mu}e_{\nu}^I+\omega_{\mu}^{~IJ}e_{\nu J}$, $H_{\mu\nu\alpha}=\del_{[\mu}B_{\nu\alpha]}$.
The internal metric is Euclidean: $\eta_{IJ}=diag[1,1,1,1]$.
In the second term above, the torsional current $J_{\mu\nu\alpha}=e_{[\mu}^I D_{\nu}(\omega)e_{\alpha] I}$, which is totally antisymmetric in its indices, has been coupled to the field strength $H_{\mu\nu\alpha}$. This coupling induces a nonvanishing torsion in this first order theory. The coupling constant $\beta$ is dimensionless. 

The Lagrangian, being a functional of three independent sets of fields $e_{\mu}^I,\omega_{\mu}^{~IJ}$ and $B_{\mu\nu}$, leads to three sets of equations of motions. We discuss these next.

\subsection{Spin connection equation}
Varying (\ref{LH}) with respect to $\omega_{\mu}^{~IJ}$, we obtain the following expression of torsion in terms of the field strength:
\begin{eqnarray*}\label{torsion}
T_{\alpha\beta}^{~~I}\equiv \frac{1}{2}D_{[\alpha}(\omega)e_{\beta]}^{I}= -\frac{\kappa}{2}~e^{\mu I}H_{\alpha\beta\mu}
\end{eqnarray*}
The spin-connection may be decomposed into a torsionless part $\omega_{\mu}^{~IJ}(e)$ and contortion $K_{\mu}^{~IJ}$ as:
\begin{eqnarray}\label{contortion}
\omega_{\mu}^{~IJ}=\omega_{\mu}^{~IJ}(e)+K_{\mu}^{~IJ}
\end{eqnarray}
Using the identity $K_{\mu\nu\alpha}=K_{\mu}^{~IJ}e_{\nu I}e_{\alpha J}=T_{\mu\alpha\nu}-T_{\nu\mu\alpha}-T_{\alpha\mu\nu}$, we obtain:
\begin{eqnarray}\label{cont}
K_{\mu\nu\alpha}=\frac{\kappa}{2}~H_{\mu\nu\alpha}
\end{eqnarray}

\subsection{$B_{\mu\nu}$ equation}
A variation of (\ref{LH}) with respect to $B_{\mu\nu}$ yields:
\begin{eqnarray*}
\del_{\mu}\left[\frac{e}{2\kappa} g^{\mu\mu'}g^{\nu\nu'}g^{\alpha\alpha'} e_{[\mu'}^I D_{\nu'}(\omega)e_{\alpha'] I}+2\beta e H^{\mu\nu\alpha}\right]=0
\end{eqnarray*}
This, upon using the connection equation, leads to: 
\begin{eqnarray}\label{B-eq}
\del_{\mu}[eH^{\mu\nu\alpha}]=0
\end{eqnarray}
Notice that this is the same equation for $B_{\mu\nu}$ as obtained in the case without torsion.
Also, due to the fact the field strength $H_{\mu\nu\alpha}$ is an exact form, there is a corresponding Bianchi identity:
\begin{eqnarray}\label{bianchi1}
\epsilon^{\mu\nu\alpha\beta}\del_{\alpha}H_{\beta\mu\nu}=0
\end{eqnarray}

%
%
\subsection{Tetrad equation:}
The only remaining set of equations of motion come from the variation of the Lagrangian with respect to the tetrad. The variations of the individual terms in (\ref{LH}) can be written as:
\begin{eqnarray}\label{eEOM}
\delta \left[\frac{1}{2}e e^{\mu}_I e^{\nu}_J R_{\mu\nu}^{IJ}(\omega)\right]&=&e \left[e^{\nu}_J R_{\rho\nu}^{KJ}(\omega)-\frac{1}{2}e_{\rho}^K R(\omega)\right]\delta e^{\rho}_K\nonumber\\
 \delta \left[eH^{\mu\nu\alpha}e_{\mu}^I D_{\nu}(\omega)e_{\alpha I}\right]&=&-e H^{\mu\nu\alpha} \left[e_{\rho}^K e^{I}_{\mu}+2 e_{\rho}^I e^{K}_{\mu}\right](D_{\nu}(\omega)e_{\alpha I})\delta e_{K}^{\rho}\nonumber\\
 &+& e H_{\rho'}^{~\nu\alpha}\left[e_{\rho}^I D_{\nu}(\omega)e_{\alpha I}+e_{\nu}^I D_{\alpha}(\omega)e_{\rho I}+e_{\alpha}^I D_{\rho}(\omega)e_{\nu I}\right]e^{(\rho'}_K \delta e^{\rho)}_K\nonumber\\
 \delta \left[eH^{\mu\nu\alpha}H_{\mu\nu\alpha}\right]&=&-e\left[e_{\rho}^K  H^{\mu\nu\alpha} H_{\mu\nu\alpha}-3 H^{~\nu\alpha}_{(\rho}H_{\mu)\nu\alpha}e^{\mu}_K  \right]\delta e^{\rho}_K 
\end{eqnarray}
which lead to:
\begin{eqnarray}\label{tetrad-eq}
R_{\rho\sigma}(\omega)-\frac{1}{2}g_{\rho\sigma}R(\omega)&=&
\kappa\left[\left(g_{\sigma\mu}e_{\nu}^I (D_{[\alpha}(\omega)e_{\rho]I})-\kappa\beta(g_{\rho\sigma}H_{\mu\nu\alpha}-6 g_{\mu\rho}H_{\sigma\nu\alpha})\right)+
\left(\rho\leftrightarrow\sigma\right)\right]H^{\mu\nu\alpha}\nonumber\\
&+&
\kappa \left[g_{\rho\sigma} e_{\mu}^{I}+g_{\mu[\sigma} e_{\rho]}^{I}\right]\left(D_{\alpha}(\omega)e_{\nu}^{I}\right)H^{\mu\nu\alpha}
\end{eqnarray}
Next, we note that using eq.(\ref{contortion}), the curvature tensor $R_{\mu\nu}^{~IJ}(\omega)$ can be expressed in terms of the curvature tensor $R_{\mu\nu}^{~IJ}(\omega(e))$ of the torsionless connection $\omega(e)$ as:
\begin{eqnarray}\label{ktorsion}
R_{\mu\nu}^{~IJ}(\omega)=R_{\mu\nu}^{~IJ}(\omega(e))+D_{[\mu}(\omega(e))K_{\nu]}^{~IJ}+K_{[\mu}^{~IL}K_{\nu]L}^{~~~J}
\end{eqnarray}
This along with eq.(\ref{cont}) allows us to rewrite the tetrad equation of motion (\ref{tetrad-eq}) as:
\begin{eqnarray}\label{e-EOM1}
R_{\rho\sigma}(\omega(e))-\frac{1}{2}g_{\rho\sigma}R(\omega(e))~=~-\kappa^2 F_a^2\left[g_{\rho\sigma} H^{\mu\nu\alpha}H_{\mu\nu\alpha}-6H_{\rho}^{~\mu\nu}
H_{\sigma\mu\nu}\right]
\end{eqnarray}
where $F_a^2=\beta-\frac{1}{8}$ is the redefined coupling constant. Also, here we have made use of the fact that the contributions from the second term in eq.(\ref{ktorsion}) vanish when the equations of motion for $\omega_{\mu}^{IJ}$ and $B_{\mu\nu}$ are used:
\begin{eqnarray}\label{k-identity}
e^{\nu}_{J}D_{[\mu}(\omega(e))K_{\nu]}^{~IJ}~=~0
\end{eqnarray}
Eq.(\ref{e-EOM1}) represents the second order theory as obtained from the first order Lagrangian (\ref{LH}). Notice that this equation is exactly the same as the Einstein equation in Giddings-Strominger theory \cite{gidd}, which admits wormhole solutions.

\section{Wormhole}
In order to discuss the wormhole configurations in this theory, we write the metric following \cite{gidd} as:
\begin{eqnarray}\label{metric}
ds^2=d\tau^2+a(\tau)^2[d\chi^2+\mathrm{sin}^2 \chi d\theta^2+\mathrm{sin}^2 \chi \mathrm{sin}^2 \theta d\phi^2]
\end{eqnarray}
Here $\tau \in [-\infty,\infty]$ is the Euclidean time and $\chi\in [0,\pi]$, $\theta\in [0,\pi]$ and $\phi\in [0,2\pi]$ are the three angles describing  a three sphere of radius $a(\tau)$. For the antisymmetric tensor field strength, we adopt the ansatz:
\begin{eqnarray}\label{H}
H^{\tau ab}=0,~H^{abc}=\frac{1}{\sqrt g}\epsilon^{abc}h(\tau,\chi,\theta,\phi)
\end{eqnarray}
where $\epsilon^{\tau abc}=\epsilon^{abc}$ is a totally antisymmetric density on the three-sphere whose indices are lowered using the induced three-metric $g_{ab}$. 

Now we proceed to find the explicit functional form of $h(\tau,\chi, \theta,\phi)$. From the equation of motion (\ref{B-eq}) for $B_{\mu\nu}$, we notice that $h$ in eq.(\ref{H}) is independent of the coordinates $\chi,\theta,\phi$:\begin{eqnarray*}h=h(\tau).\end{eqnarray*} The Bianchi identity $\del_{[\tau}H_{\chi \theta\phi]}=0$ fixes the $a(\tau)$ dependence of $h(\tau)$ as:
\begin{eqnarray}\label{h}
h(\tau)=\frac{\kappa Q}{3!~a^3(\tau)}
\end{eqnarray}
where, $Q$ is a dimensionless constant. For an appropriate normalization, $Q$ turns out to be the axion charge, which is given by the integral of the field strength over any three-surface:
\begin{eqnarray}\label{charge-int}
\int d^3 x ~\epsilon^{abc}H_{abc}=2\pi^2 \kappa Q
\end{eqnarray}
The  above ansatz for $H_{\mu\nu\alpha}$, when inserted into the tetrad equations of motion (\ref{e-EOM1}), leads to:
\begin{eqnarray}\label{wh}
\dot{a}^2(\tau) =1-\frac{\kappa^4 F_a^2 Q^2}{18 a^4(\tau)}
\end{eqnarray}
This equation describes the Giddings-Strominger wormhole configuration. For the minimum value $a_0=18^{-\frac{1}{4}}\kappa (F_a Q)^\frac{1}{2}$ of the scale factor, $\dot{a}(\tau)=0$. This corresponds to the size of the throat of the wormhole. The explicit form of the solution, as discussed in \cite{gidd}, shows that this configuration interpolates between two asymptotically flat surfaces which are topologically $R^3$ at any instant of time. The wormhole at any fixed time represents an incontractible three sphere. Starting with some large radius at $\tau \rightarrow -\infty$, the wormhole attains the minimum size $a_0$ at $\tau=0$ and then again reaches the maximal radius at $\tau \rightarrow \infty$. On the other hand, the half-wormhole represents a tunneling configuration for an $R^3$ geometry at $\tau =-\infty$ to $R^3+S^3$ at $\tau=0$ \cite{gidd}. 

In the next section, we demonstrate that the first order gravity theory allows a new interpretation of these instanton configurations as torsional pseudoparticles (Nieh-Yan instantons).

\section{Torsion instanton}

\subsection{Nieh-Yan Topological charge:}
The semi-wormhole configuration in the first-order theory is associated with a nontrivial torsion. The associated Nieh-Yan index, which is the only topological invariant associated with torsion, is defined as \cite{nieh}:
\begin{eqnarray}\label{nieh-yan}
N_{NY}~&=&~\frac{1}{2\pi^2 \kappa^2}\int_{M^4} d^4 x~ \epsilon^{\mu\nu\alpha\beta}\left[e_{\mu}^{I}e_{\nu}^{J}
R_{\alpha\beta}^{~IJ}(\omega)~-~2\left(D_\mu(\omega)e_{\nu}^I\right)\left(D_\alpha(\omega)e_{\beta}^I\right)\right]\end{eqnarray}
The integrand above is a total divergence:
\begin{eqnarray}
N_{NY}~=~-\frac{1}{\pi^2 \kappa^2}\int_{M^4} d^4 x~ \del_{\mu}\left[\epsilon^{\mu\nu\alpha\beta}e_{\nu}^I D_\alpha(\omega)e_{\beta}^I\right]
\end{eqnarray}
For a four manifold $M^4$ with a compact boundary $\del M$, this reduces to:
\begin{eqnarray*}
N_{NY}~&=&~-\frac{1}{\pi^2 \kappa^2}\int_{\del M} d^3 x ~\left[\epsilon^{abc}e_{a}^I D_b(\omega)e_{c}^I\right]
\end{eqnarray*}
For the Giddings-Strominger configuration, the only compact boundaries of the four manifold are the baby universes which are topologically $S^3$. Using the decomposition of the spin-connection and the expression for $K_{\mu}^{~IJ}$, the Nieh-Yan number of the semi-wormhole thus becomes:
\begin{eqnarray}\label{nycharge}
N_{NY}~=~\frac{1}{\pi^2 \kappa^2}\int_{S^3} d^3 x ~[\epsilon^{abc}K_{abc}]~=~\frac{1}{2\pi^2 \kappa}\int_{S^3} d^3 x ~\epsilon^{abc}H_{abc}~=~ Q
\end{eqnarray}
This implies an exact equality between the Nieh-Yan index of the instanton and axion charge $Q$ carried away by the baby universe. 
It is well known that the Nieh-Yan index of a compact manifold can be expressed as the difference of $SO(5)$ and $SO(4)$ Pontryagin numbers of the same manifold \cite{zanelli}. Thus, for $\del M=S^3$, the winding number is nothing but a combination of the homotopy indices associated with $\pi_3[SO(5)]$ and $\pi_3 [SO(4)]$, which are known to be integers:
 \begin{eqnarray}\label{quant}
N_{NY}|_{S^3}=\pi_3[SO(5)]+\pi_3[SO(4)]=Z+(Z+Z)
\end{eqnarray}
Thus, if the initial configuration, which is topologically $R^3$, has zero axion charge, instanton effects would lead to a final configuration with a nonvanishing axion charge $-Q$, where $Q=\sum_{i}Q_i$ is the sum of charges carried away by all the baby universes. 

It is important to note that since the Nieh-Yan number is an integer, the axion charge as related to it through eq.(\ref{nycharge}) is quantized. A similar quantization property of axion charge has been noticed in the context of oriented bosonic strings coupled to the antisymmetric tensor field $B_{\mu\nu}$ \cite{rohm}. However, in that case, the integers associated with the axion charge have a different origin, namely, the nontrivial third cohomology group (with integer coefficients) $H^3(S^3,Z)$ of the three-sphere. 


\subsection{Action of instanton:}
In order to compute the action, we make use of the connection equation (\ref{cont}) and identities (\ref{ktorsion}) and (\ref{k-identity}) to write:
\begin{eqnarray}\label{k-identity1}
R(\omega)=R(\omega(e))-\frac{\kappa^2}{4}H^{\rho\lambda\sigma}H_{\rho\lambda\sigma}
\end{eqnarray}
where $H_{\mu\nu\alpha}$ are given by eq.(\ref{H}) for the wormhole solution. The above equation along with (\ref{cont}) and (\ref{e-EOM1}) allow us to rewrite the Lagrangian density (\ref{LH}) as:
\begin{eqnarray*}
L(e,B)&=&-\frac{1}{2\kappa^2}eR(\omega(e))~+~eF_a^2 H^{\mu\nu\rho}H_{\mu\nu\rho}~=~12e F_a^2 h^2(\tau) 
\end{eqnarray*}
The resulting action for a semiwormhole (instanton) reads:
\begin{eqnarray}\label{action}
S =\int d^4 x~ L(e,B)
&=&  \frac{2\pi^2}{3}\kappa^2 F_a^2 Q^2 \int_{\tau_0}^{\infty}\frac{d\tau}{a^3(\tau)}\nonumber\\
&=& \frac{2\pi^2}{3}\kappa^2 F_{a}^{2} Q^2 \int^{\infty}_{a_0} da(\tau)~a^{-3}(\tau)\left[1-\frac{a_0^4}{a^4(\tau)}\right]^{-\frac{1}{2}}\nonumber\\
&=& \frac{~\pi^3}{\sqrt{2}}F_a Q
\end{eqnarray}
where in the second line we have used eq.(\ref{wh}).
Importantly, the action for the instanton is finite, and is expected to contribute nontrivially in the functional integral for this theory. 

\section{$\eta$-vacuum}
Half-wormholes induce quantum tunneling between classical vacuum states of different Nieh-Yan numbers $N_{NY}$. For each such state, this number is given by the difference of the total positive and negative charges carried by the baby universes in that state. This implies that the quantum vacuum is a linear superposition of the classical vacua:
\begin{eqnarray}
\ket{\eta}=\sum_{N_{NY}}e^{i\eta N_{NY}}\ket{N_{NY}} \label{qvac}
\end{eqnarray}
The transition amplitude between states of different Nieh-Yan numbers can then be calculated in the same manner as in the gauge theories \cite{raja}. 

 We adopt a dilute gas approximation where the instantons (anti-instantons), which carry a Nieh-Yan (axion) charge $+1$ ($-1)$, are widely separated in the interpolating four geometry. We also assume each of the baby universes in the initial and final states to be of charge $\pm 1$. Let the initial vacuum state $\ket{N^i_{NY}=N^{i}_{+}-N^{i}_{-}}$ have $N^i_{\pm}$ universes of charges  $\pm 1$ at some early time and the final vacuum state 
 $\ket{N^f_{NY}=N^{f}_{+}-N^{f}_{-}}$  have $N^f_\pm  $ universes of charge $\pm 1$ at   distant future. As time progresses, we include the contribution of $n_+$ instantonic configurations, each of these creating a baby universe of Nieh-Yan charge $+1$ in the final state as depicted in Fig 1(a). In addition, we can have $n_-$ configurations, each annihilating a baby universe of charge $-1$ in the initial state as depicted in Fig 1(b). Similarly, we can have ${\bar n}_{\pm}$ half-wormholes, each annihilating a baby universe of Nieh-Yan charge $+1$ from the initial state or creating a baby universe of charge $-1$ in the final state as depicted in Fig 1 (c) and (d), respectively.
\begin{figure}\begin{center}
\includegraphics[height=5cm]{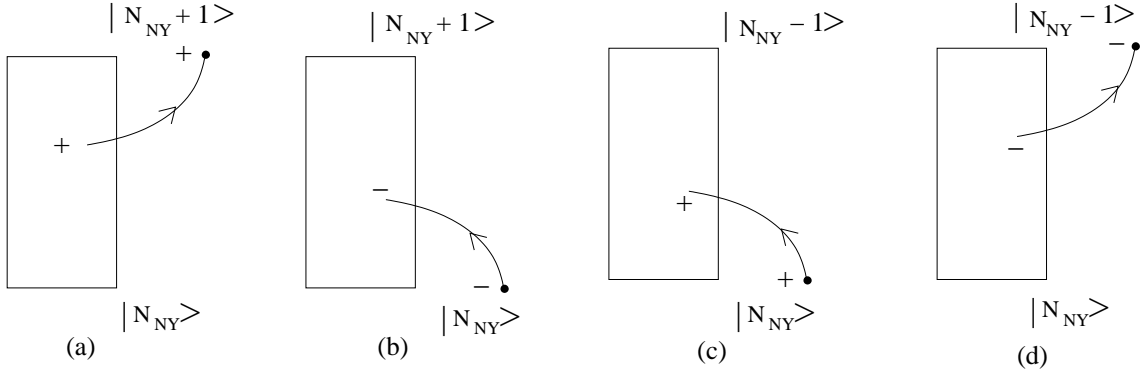}
\caption{Tunneling configurations}
\end{center}
\end{figure}
However, among these, the   diagrams (a) and (b) represent equivalent configurations, since they describe the same tunneling process, i.e. 
one that induces a change in the Nieh-Yan charge by $+1$ between the 
initial and final states. 
 Similarly,   diagrams (c) and (d) represent the same amplitude
 inducing a change of the Nieh-Yan charge by $-1$.  Thus, the $n_{+}+\bar{n}_{-}=N$ configurations should be treated as indistinguishable, and so should be the $n_{-}+\bar{n}_{+}=\bar{N}$ semi-wormholes.   The difference of the initial and final Nieh-Yan numbers is fixed:
  \begin{eqnarray*}
 N_{NY}^{f}-N_{NY}^{i}=(n_{+}-\bar{n}_{+})-(n_{-}-\bar{n}_{-})=N-\bar{N}
 \end{eqnarray*}
In the dilute gas approximation, each configuration contributes a factor proportional to $e^{-S}KVT$ to the path-integral where $S$ is the action (\ref{action}) corresponding to an instanton or anti-instanton and $K$ represents the contribution from quantum fluctuations around the wormhole solution. The factor $VT$ is the volume of the interpolating four manifold which comes due to an integration over the location of each instanton (anti-instanton). Here $K$ is assumed to be real \cite{gidd1} and it is independent of $VT$ in the large volume limit. 
The cumulative contribution of all the instantons and anti-instantons of charge $\pm 1$ to the path-integral can be written as:
\begin{eqnarray}\label{amp}
\braket{N^f_{NY}|e^{-FT}|N^i_{NY}} &=& A~\sum_{N=0}^{\infty}\sum_{\bar{N}=0}^{\infty}~\frac{1}{N!~\bar{N}!}~\left(e^{-S}KVT\right)^{N+\bar{N}}[\delta_{(N^f_{NY} -N^i_{NY}),(N-\bar{N})}] 
\nonumber\\
&=& A \int_{0}^{2\pi} d\eta ~\sum_{N=0}^{\infty}\sum_{\bar{N}=0}^{\infty}~e^{i\eta(N^i_{NY}-N^f_{NY}+N-\bar{N})} \frac{\left(e^{-S} KVT\right)^N}{N!} 
\frac{\left(e^{-S} KVT\right)^{\bar{N}}}{\bar{N}!} \nonumber\\ 
&=& A \int_{0}^{2\pi} d\eta ~e^{i\eta(N^i_{NY}-N^f_{NY})}~ \mathrm{exp}\left[2e^{-S}KVT \mathrm{cos}\eta\right]
\end{eqnarray}
Here, $A$ is a normalization factor and $F$ has been defined to be the formal analogue of the vacuum energy of the system, although its precise relation to the energy in gravity theory is not clear \cite{stro}. 
This  allows us to write the instanton and anti-instanton contributions to the transition amplitude  
for the nonperturbative quantum $\eta$-vacua of Eqn. (\ref{qvac}) as:
\begin{eqnarray}\label{etaamp}
\braket{\eta'|~ e^{-FT} ~|\eta} = A ~\delta(\eta - \eta') ~\exp\left[2e^{-S}KVT \mathrm{cos}\eta\right]
\end{eqnarray}
The main thrust of the above result is that the vacuum energy receives an
 $\eta$-dependent modification of the size $ F _{\eta}=- 2e^{-S}KV cos\eta$ due to tunneling effects. This vacuum energy has been evaluated earlier in \cite{gidd} for the ground state constructed as a linear combination of perturbative vacuum states with different baby universe number.

The fact that there exists a non-perturbative $\eta$-vacuum can be captured, like in QCD, through an effective Euclidean Lagrangian containing  Nieh-Yan topological
density with coefficient $\eta$ as an additional term:
\begin{eqnarray}\label{l-eff}
L_{eff}=L+i\eta I_{NY}
\end{eqnarray} 
where $L$ is the original Lagrangian as defined in (\ref{LH}). It is well known that in first order gravity Lagrangian,  addition of Nieh-Yan density does not change the classical dynamics. The topological coupling constant $\eta$ multiplying this term is identified \cite{date,kaul} as the inverse of the Barbero-Immirzi parameter of Loop Quantum Gravity. To sum it up, we have found a $\eta$-vacuum in first order gravity, which is a result of tunneling between perturbative vacua labelled by different Nieh-Yan indices. This is a clear realization of the suggestion made earlier \cite{gambini} that Barbero-Immirzi parameter should show up through a rich vacuum structure in quantum gravity \cite{date,kaul,sengupta,sengupta1,mercuri}.

 Since Nieh-Yan density is $P$ and $T$ odd, the $\eta$ parameter is a   
 quantum  coupling constant violating these symmetries. 
Like in the case of $\theta$-vacuum of QCD \cite{raja}, let us consider the expectation value of the Nieh-Yan index in the $\eta$-vacuum, which, according to equations (\ref{amp}) and (\ref{l-eff}) is given by: 
\begin{eqnarray}\label{anomaly}
\bra{\eta}\int d^4 x~I_{NY}\ket{\eta}
~=~ \frac{i}{VT}\frac{d}{d\eta}\left(\mathrm{ln} ~e^{-F_{\eta}T}\right)~=~ -2i K e^{-S} \mathrm{sin}\eta
\end{eqnarray}
where  $K$ is a calculable quantity leaving $\eta$ as the only free parameter in this expression. 
We emphasize that the instanton effects reflected by this equation are insensitive to the specific details of the origin of torsion.   For example,
as fermions couple to contortion, the divergence of the axial current would get a contribution as in eq.(\ref{anomaly}) for any configuration with a nontrivial Nieh-Yan number. This is because the axial anomaly is known to be related to the expectation value of the Nieh-Yan index \cite{zanelli}. Thus, the electric dipole moment of neutron can develop a dependence on the Barbero-Immirzi parameter $\eta$ (see ref.\cite{eguchi,hojman} for earlier discussions on parity violating effects in gravity). 

We have already discussed how the Nieh-Yan number can be expressed in terms of the contortion field $K_{\mu\nu\alpha}$ (eq.(\ref{nycharge})).  
We conclude this
section by discussing the nature of large gauge transformations which induce
change of Nieh-Yan number of a perturbative state.
 \vspace{.2cm}\\
{\bf Large gauge transformations:}
\vspace{.1cm}\\
Let us consider the following transformation of $K_{\mu\nu\alpha}$:
\begin{eqnarray}\label{gauge}
\delta K_{\mu\nu\alpha}=\frac{1}{3!g}\epsilon_{\mu\nu\alpha\beta}\zeta^{\beta}
\end{eqnarray}
where $g$ is the determinant of the four-metric $g_{\mu\nu}$. Also, here we have introduced a `gauge vector' $\zeta^{\beta}:=\left(\frac{\Lambda(\tau)}{2\pi^2},0,0,0\right)$ where 
 $\Lambda(\tau)$ is a smooth function of $\tau$ such that $\Lambda(-\infty)=0$ and $\Lambda(\infty)=1$. Since the gauge vector is non-vanishing at infinity, it represents a `large' gauge transformation of the contortion field. To be specific, we can choose\cite{raja}: 
\begin{eqnarray*}
\Lambda(\tau)=\frac{1}{2}(1+\mathrm{tanh}\tau)
\end{eqnarray*}
Hence, under (\ref{gauge}), the change in the Nieh-Yan   number  is given by:
\begin{eqnarray}\label{index}
\delta \int d^4 x~ \epsilon^{\mu\nu\alpha\beta}\del_{\mu}K_{\nu\alpha\beta}~=~
\int d^4 x~ \del_{\mu}\zeta^{\mu}~=~\Lambda(\tau)|_{-\infty}^{+\infty}~=~1
\end{eqnarray}

Our analysis in this paper is based on a first order theory of  
gravity where the source of torsion is an axion. However, there can
be other matter induced sources of torsion and also those of
a geometric origin \cite{regge,tseytlin,zanelli}. Our discussion above generically applies
to the $\eta$-vacuum generated by all such sources of torsion.

%

\section{Concluding remarks}

 Here we provide a new interpretation of the instanton number of the Giddings-Strominger (semi) wormhole in terms of the Nieh-Yan topological invariant. The nonperturbative vacuum structure arising due to tunneling between states of different Nieh-Yan numbers is found to be characterised by the topological coupling constant $\eta$, which can be identified with the Barbero-Immirzi parameter of Loop Quantum Gravity. Thus, $\eta$ emerges as an exact analogue of the $\theta$ angle in gauge theories. That the Barbero-Immirzi parameter has such a topological origin was anticipated earlier in \cite{gambini} and demonstrated in \cite{date} in the context of classical canonical gravity. Subsequently, this idea has been explored in several places \cite{kaul,sengupta,mercuri,sengupta1}. However, there has been no analysis or demonstration of possible nonperturbative effects that could be induced by Nieh-Yan instantons in the quantum theory of gravity. Our work here provides a concrete example of an instanton which can be used to extract the quantum physics associated with the Nieh-Yan invariant, or equivalently, with the Barbero-Immirzi parameter.

We also find that the axion charge is given in terms of the homotopy integers associated with the Nieh-Yan invariant, and hence is quantized. There is a similar quantization condition in Bosonic string theory with an axion coupling. Whether these two constraints are independent or are related is to be understood.

Finally, we emphasize that although our analysis here was based on the axionic wormholes, the essential features of the $\eta$-vacuum as unravelled here are expected to go through for any other configuration corresponding to a nontrivial torsion in the first order theory. This can be appreciated by noticing that the discussions on Nieh-Yan number and large gauge transformations have been formulated entirely in terms of the contortion field $K_{\mu\nu\alpha}$, which can be generated either by some other matter-coupling or by a nontrivial geometry. 
%
\acknowledgments 
R.K.K gratefully acknowledges the hospitality during June
and July, 2014 at the Center for High Energy Physics, Indian
Institute of Science, Bangalore where part of the work reported
here was done. He also acknowledges the support of Department
of Science and Technology, Government of India, through a J.C.
Bose National Fellowship.

S.S. acknowledges many insightful discussions with Sayan Kar during December, 2013 at IIT Kharagpur where this work was initiated. Conversations with Joseph Samuel and Madhavan Varadarajan have also been rewarding, and so have been the comments of Fernando Barbero, Martin Bojowald, Miguel Campiglia and Siddhartha Lal.

\end{document}